\documentclass[12pt]{article}
\usepackage{axodraw}

\newcommand{\beq}{\begin{equation}}
\newcommand{\eeq}{\end{equation}}
\newcommand{\beqa}{\begin{eqnarray}}
\newcommand{\eeqa}{\end{eqnarray}}
\newcommand{\beqar}{\begin{eqnarray*}}
\newcommand{\eeqar}{\end{eqnarray*}}

\newcommand{\al}{\alpha}
\newcommand{\be}{\beta}

\def\spa          {\ \ \ }
\def\non          {\nonumber}
\def\ha           {\mbox{$\frac{1}{2}$}}

\def\d        {\mbox{d}}
\def\spa          {\ \ \ }
\def\mand         {\spa\mbox{and}\spa}

\def\Tr           {\mbox{\rm Tr}\,}
\def\STr          {\mbox{\rm STr}\,}
\def\Str          {\mbox{\rm Str}\,}

\def\cd           {{\cdot}}
\def\ran          {\rangle}
\def\lan          {\langle}
\def\fsH	{H\!\!\!\!/\,}
\newcommand{\del}{\delta}

\newcommand{\eps}{\epsilon}
\newcommand{\ga}{\gamma}
\newcommand{\Ga}{\Gamma}

\newcommand{\inn}{\!\cdot\!}
\newcommand{\h}{\eta}

\newcommand{\lam}{\lambda}

\newcommand{\z}{\zeta}

\newcommand{\eg}{{\it e.g.,}\ }
\newcommand{\ie}{{\it i.e.,}\ }
\newcommand{\labell}[1]{\label{#1}} 
\newcommand{\reef}[1]{(\ref{#1})}
\newcommand\prt{\partial}

\newcommand\cL{{\cal L}}
\newcommand\cD{{\cal D}}

\newcommand\bz{\bar{z}}

\newcommand\cT{T}

\begin{document}
\baselineskip 18pt%
\begin{titlepage}
\vspace*{1mm}%
\hfill%
\vspace*{15mm}%

\centerline{{\Large {\bf On effective actions of  non-BPS branes  }}}\vspace*{3mm} \centerline{{\Large {\bf  
and their  higher derivative corrections  
 }}}
\vspace*{5mm}
\begin{center}
{ Mohammad R. Garousi }%

\vspace*{0.8cm}{ {Department of Physics, Ferdowsi university of Mashhad,\\
P.O. Box 1436, Mashhad, Iran}}\\
{\it {and}}\\
{  School of Physics, 
 Institute for research in fundamental sciences (IPM), \\
 P.O. Box 19395-5531, Tehran, Iran. }\\
\vspace*{1.5cm}
\end{center}

\begin{center}{\bf Abstract}\end{center}
\begin{quote}
By calculating  various disk level S-matrix elements and studying in details their momentum expansions, we have extracted   some of the couplings in tachyon DBI action and Wess-Zumino terms of the non-BPS branes, and their higher derivative corrections. In particular, we have  found  that there is exact consistency between field theory and string theory tachyon pole of S-matrix element of one RR and three tachyons  provided that  one takes into account  the fact that   the tachyon vertex operator in 0 picture to be along   the Pauli matrix $\sigma_1$  whereas the  tachyon  in -1 picture to be  along the $\sigma_2$ direction. This internal CP factors should be included in the   tachyon DBI part of the effective action.

\end{quote}
\end{titlepage}
\section{Introduction}
Study of unstable objects in string theory  might shed new light
in understanding properties of string theory in time-dependent
backgrounds \cite{Gutperle:2002ai,Sen:2002in,Sen:2002an,Sen:2002vv,Lambert:2003zr,Sen:2004nf}. The source of instability   is appearance
of some tachyon  modes  in the spectrum of these 
objects. It  then makes sense to study them in a field
theory which includes the massless and tachyon fields. In this regard, it has been
shown by A. Sen that an effective action of  Born-Infeld type
proposed in \cite{Sen:1999md,Garousi:2000tr,Bergshoeff:2000dq,Kluson:2000iy} can capture many properties
of the decay of non-BPS D$_p$-branes in string theory
\cite{Sen:2002in,Sen:2002an}. 

The non-BPS D$_p$-branes of type IIA(B) string theory are defined by projecting D$_p$-brane-anti-D$_p$-brane of type IIB(A) with $(-1)^{F_L}$ where $F_L$ denotes the contribution to the space-time fermion number from the left-moving sector of the string world-sheet \cite{Sen:1999mg}. The open strings of the brane-anti-brane can be labeled by the ${\it external}$  $2\times 2$ Chan-Paton factors:
\beqa
(a):\,\pmatrix{0&0\cr 
0&1},\,\,\,
(b):\,\pmatrix{1&0\cr 
0&0},\,\,\, (c):\,\pmatrix{0&0\cr 
1&0},\,\,\, (d):\,\pmatrix{0&1\cr 
0&0}\labell{M121} \nonumber\eeqa   
The massless states carry CP factor (a), (b), and the tachyons carry (c) and (d) factors. The projection operator $(-1)^{F_L}$ has no effect on the world-sheet fields, however, using the fact that it exchanges brane with anti-brane, one observes that its effect on the CP matrix $\Lambda$ is the following: 
\beqa
\Lambda\rightarrow \sigma_1\Lambda (\sigma_1)^{-1}, \nonumber\eeqa
where $\sigma_1$  is the Pauli matrix.  The states with CP factors $I$ and $\sigma_1$ are survived. The massless fields then carry the ${\it internal}$ CP factor $I$ and the real tachyon of non-BPS D-brane carries the CP factor $\sigma_1$. The vertex operators corresponding  to the above states appear in S-matrix elements in different pictures. We assign the above internal CP factors  to the vertex operators in the 0 picture. Using the observation made in \cite{DeSmet:2000je} that the picture changing operator on a non-BPS brane naturally comes with $\sigma_3$, one observes that in -1 picture, the internal CP factor of massless states is $\sigma_3$ and the CP factor of tachyon is $\sigma_2$.

Another interesting object in string theory is SD$_p$-brane. In field theory, as the kink solution of the non-BPS D-brane effective action gives BPS D-brane, the kink solution in the time direction gives S-brane. In  string theory, it  is defined  as an object on which open strings with Dirichlet boundary condition on the first $9-p$ directions $X^a$, $a=0,1,\cdots, 8-p$ and  Neumann boundary condition on the last $p+1$ coordinates $X^i$, $i=9-p,\cdots, 9$ end \cite{Gutperle:2002ai}. The SD$_p$-branes in Type II theory carry the RR charges, and the non-BPS SD$_p$-branes  of type IIA(B) string theory are defined by projecting SD$_p$-brane-anti-SD$_p$-brane of type IIB(A) with $(-1)^{F_L}$. 

The effective action of a non-BPS brane/brane-anti-brane has two parts, the part which is an extension of DBI action and the Wess-Zumino part, \ie 
\beqa
S_{non-BPS}&=&S_{DBI}+S_{WZ}\labell{nonbps}\nonumber\eeqa
The  WZ 
term describing the  coupling of RR field to the gauge field of brane-anti-brane  is given by~\cite{Douglas:1995bn,Li:1995pq}
\beqa
S = \mu_p\int_{\Sigma_{(p+1)}} C \wedge \Tr \left(e^{i2\pi\alpha'F^{(1)}}-e^{i2\pi\alpha'F^{(2)}}\right)\ ,
\labell{eqn.wz}
\eeqa
where $\Sigma_{(p+1)}$ is the world volume and $\mu_p$ is the RR charge of the branes. In above equation, $C$ is a formal sum of the RR potentials $C=\sum_{m=p+1}(-i)^{\frac{p+1-m}{2}}C_m$. Note that the factors of $i$ disappear in  each term of \reef{eqn.wz}.
The inclusion of the tachyon fields into this action has been proposed in \cite{Kennedy:1999nn,Kraus:2000nj,Takayanagi:2000rz} using the superconnection of noncommutative geometry ~\cite{quil,berl,Roepstorff:1998vh}\beqa
S_{WZ}&=&\mu_p \int_{\Sigma_{(p+1)}} C \wedge \STr e^{i2\pi\alpha'\cal F}\labell{WZ}\eeqa 
where the  ``supertrace'' is defined as:
\begin{displaymath}
\STr \left( \begin{array}{cc} A&B\\C&D \end{array} \right)
= \Tr A - \Tr D \ .
\non\end{displaymath} and the curvature of the superconnection is defined as:
\beqa {\cal F}&=&d{\cal A}-i{\cal A}\wedge\cal A\nonumber\eeqa
the superconnection is \begin{displaymath}
i{\cal A} = \left(
\begin{array}{cc}
  iA^{(1)} & \beta T^* \\ \beta T &   iA^{(2)} 
\end{array}
\right) \ ,
\non\end{displaymath}
where $\beta$ is a normalization constant with dimension $1/\sqrt{\alpha'}$.  If one uses the standard non-abelian kinetic term for  the dynamics  of the tachyon  field then the normalization of tachyon in the WZ action  \reef{WZ} has to be  \cite{Garousi:2007fk}
\beqa
\beta&=&\frac{1}{\pi} \sqrt{\frac{2\ln(2)}{\alpha'}}\labell{norm0}\eeqa

The WZ terms of the non-BPS branes  can also be written   in terms of superconnection \cite{Kraus:2000nj,Takayanagi:2000rz},
\beqa
S_{WZ}&=&\mu_p' \int_{\Sigma_{(p+1)}} C \wedge \Str e^{i2\pi\alpha'\cal F}\labell{WZ'}\eeqa 
where the ``supertrace'' is now 
\begin{displaymath}
\Str \left( \begin{array}{cc} A&B\\C&D \end{array} \right)
= \Tr B + \Tr C \ .
\non\end{displaymath}  and the superconnection is
 \begin{displaymath}
i{\cal A} = \left(
\begin{array}{cc}
  iA & \beta' T \\ \beta' T &   iA 
\end{array}
\right) \ ,
\non\end{displaymath}
where  $\beta'$ is a new normalization constant with dimension $1/\sqrt{\alpha'}$.  Using the multiplication rule of the supermatrices \cite{Kraus:2000nj}
\beqa
 \left
( \begin{array}{cc}A&B\\C&D\end{array} 
\right)\cdot \left( \begin{array}{cc}A'&B'\\C'&D'\end{array} 
\right)\,=\,\left( \begin{array}{cc}AA'+(-)^{c'}BC'&AB'+(-)^{d'}BD'\\DC'+(-)^{a'}CA'&DD'+(-)^{b'}CB'\end{array} 
\right)\nonumber\eeqa where $x'$ is 0 if $X$ is an even form or 1 if $X$ is an odd form, one finds that the curvature   of the superconnection is 
\begin{displaymath}
i{\cal F} = \left(
\begin{array}{cc}
iF -\beta'^2 T^2 & \beta' DT \\
\beta' DT & iF -\beta'^2T^2 
\end{array}
\right) \ ,
\non\end{displaymath}
where $F=\frac{1}{2}F_{ab}dx^{a}\wedge dx^{b}$ and $DT=(\partial_a T-i[A_{a},T])dx^{a}$. In  equation \reef{WZ'}, 
 $C$ is a formal sum of the RR potentials  $C=\sum_{m=p}(-i)^{\frac{p-m}{2}}C_m$.
Using the expansion for the exponential term in the WZ action \reef{WZ'}, one finds 
the following terms:
\beqa
\mu_p'(2\pi\alpha')C\wedge \Str i{\cal F}&\!\!\!\!=\!\!\!&2\beta'\mu_p' (2\pi\alpha')\Tr\left(C_{p}\wedge DT\right)\labell{exp2}\\
\frac{\mu_p'}{2!}(2\pi\alpha')^2C\wedge \Str i{\cal F}\wedge i{\cal F}&\!\!\!\!=\!\!\!\!&2\beta'\mu_p'(2\pi\alpha')^2\Tr\left(-\beta'^2T^2C_{p}\wedge DT  +C_{p-2}\wedge DT\wedge F\right)\nonumber\\
\frac{\mu_p'}{3!}(2\pi\alpha')^3C\wedge \Str i{\cal F}\wedge i{\cal F}\wedge i{\cal F}&\!\!\!\!=\!\!\!\!&
\beta'\mu_p'(2\pi\alpha')^3\Tr\left(\beta'^4T^4C_{p}\wedge DT+C_{p-4}\wedge F\wedge F\wedge DT\right.\nonumber\\
&&\left.+C_{p-2}\wedge\left(2\beta'^2T^2F\wedge DT+i\frac{\beta'^2}{3}DT\wedge DT\wedge DT\right)\right)\nonumber
\eeqa
where the trace is over the $U(N)$ matrices. 

We  shall confirm  the above couplings with the S-matrix method, and find their higher derivative corrections. The S-matrix method can not be used to confirm the above WZ couplings in the non-BPS D-branes  due to some kinematic reason \cite{Billo:1999tv}. However, there is no such kinematic obstacle in the  non-BPS SD-brane case. To see this consider the coupling $C\wedge dT$. In the momentum space it is proportional to the tachyon momentum $k$.
The conservation of momentum along the world-volume implies that $k^2=p_ip^i$ where $p_i$ is the RR momentum along the world-volume.  Using  the on-shell condition $m^2=-k^2=-1/4$ for tachyon\footnote{Our convention sets $\alpha'=2$.}  and $p_ip^i=-p_ap^a$ for the massless RR field, one finds that the RR momentum along the brane must be non-zero, \ie $p_ip^i=1/4$ which is consistent with $p_ip^i=-p_ap^a$  as $p_ap^a$ can be negative for non-BPS SD-brane. For the non-BPS D-brane, on the other hand,  the closed string on-shell condition is $p_ip^i=-p_ap^a<0$ which is not consistent with $p_ip^i=1/4$. Hence, the S-matrix method gives only the WZ couplings of non-BPS SD-branes. 

An outline of the rest of paper is as follows. In the next section,  to fix our notation we calculate  the S-matrix element of one RR and one tachyon  \cite{Kennedy:1999nn}. This gives the coupling $C_p\wedge DT$ in \reef{WZ'}. In  section 3, we calculate the S-matrix element of one RR, one tachyon   and one gauge field. The momentum expansion of this amplitude gives the coupling $C_{p-2}\wedge DT\wedge F$ in \reef{WZ'} and its higher derivative corrections. In section 4, we calculate the S-matrix element of one RR and three tachyons and find its  momentum expansion. This amplitude  has two parts. The first part which involves $C_{p-2}$ has massless pole and contact terms. The massless poles are reproduced by the higher derivative couplings of $C_{p-2}\wedge DT\wedge F$   and the standard non-abelian kinetic term of tachyon. The contact terms give coupling $C_{p-2}\wedge DT\wedge DT\wedge DT$ in \reef{WZ'} and its higher derivative corrections. The second part which involve $C_p$ has tachyon pole and contact terms. The contact terms give the coupling $T^2C_p\wedge DT $ and its higher derivative corrections.  To produce the tachyon poles, one needs to know the couplings of four tachyons.  In section 4.2.1, using the fact that at disk level two tachyons must carry $\sigma_1$ and the other two tachyons must carry the internal CP factor $\sigma_2$, we calculate  the S-matrix element of four tachyons from which  we will find a bunch of  higher derivative couplings of  four tachyons. Using these couplings, we will show that the infinite tower of the tachyon poles of the S-matrix element of one RR and three tachyons are reproduced exactly by field theory.  We discuss our results in section 5.


\section{The $T-C$ amplitude}

 The two point amplitude between one RR and one  tachyon is given by the following correlation functions:
\begin{eqnarray}
{\cal A}^{T,RR} & \sim & \int dyd^2z
 \lan 
V_{T}^{(-1)}(y)
V_{RR}^{(-1)}(z,\bar{z})\ran\labell{cor10}\eeqa
where the vertex operators are
\beqa
V_{T}^{(-1)}(y) &=& e^{-\phi(y)} e^{2ik\cd X(y)}\lam\otimes\sigma_2\labell{vertex1}\\
V_{RR}^{(-1)}(z,\bar{z})&=&(P_{-}\fsH_{(n)}M_p)^{\al\be}e^{-\phi(z)/2} S_{\al}(z)e^{ip\cd X(z)}e^{-\phi(\bar{z})/2} S_{\be}(\bar{z}) e^{ip\cd D \cd X(\bar{z})}\otimes\sigma_3\sigma_1\nonumber
\eeqa
where $k^i$ is the tachyon momentum and $\lam$ in the tachyon vertex operator is matrix in the $U(N)$ group.  The RR vertex operator of  brane-anti-brane system should carry  the internal Pauli matrix  $\sigma_3$. To see this, we note that  the $\sigma$-factor of the S-matrix element of one $\sigma_1$-type tachyon , one $\sigma_2$-type tachyon and one RR vertex operator  of brane-anti-brane is non-zero which is consistent with the WZ terms of brane-anti-brane system \cite{Garousi:2007fk}. Without the $\sigma_3$ factor the $\sigma$-factor of the S-matrix element is  $\Tr(\sigma_1\sigma_2)=0$  which is not consistent with the WZ couplings. For non-BPS branes, there should be  an extra factor of $\sigma_1$ in  the RR vertex operator.  As argued in \cite{Sen:1999mg}, since the RR state of non-BPS brane appears in the twisted sector when one regards type IIB as type IIA modded out by $(-1)^{F_L}$, there is a cut extending from RR vertex operator at the center of disk all the way to the boundary of the disk. At the point where the cut hits the boundary one needs to insert an extra factor of $\sigma_1$.\footnote{ In the conventions \cite{Sen:1999mg},  the RR field which is in -2 picture comes with a single Pauli matrix $\sigma_1$,  in our conventions, it is in -1 picture hence it comes with $\sigma_3\sigma_1$.} The  CP factor in the S-matrix element \reef{cor10} is  $\Tr(\lam)\otimes\Tr(\sigma_2\sigma_3\sigma_1)$.

The projector in the RR vertex operator is $P_{-} = \ha (1-\ga^{11})$  and
\begin{displaymath}
\fsH_{(n)} = \frac{a
_n}{n!}H_{\mu_{1}\ldots\mu_{n}}\ga^{\mu_{1}}\ldots
\ga^{\mu_{n}}
\ ,
\non\end{displaymath}
where $n=2,4$ for type IIA and $n=1,3,5$ for type IIB. $a_n=i$ for IIA and $a_n=1$ for IIB theory. The spinorial
indices are raised with the charge conjugation matrix, \eg
$(P_{-}\fsH_{(n)})^{\al\be} =
C^{\al\del}(P_{-}\fsH_{(n)})_{\del}{}^{\be}$ (further conventions and
notations for spinors can be found in appendix~B of~\cite{Garousi:1996ad}).  The
RR bosons are massless so $p^{2}=0$ and for the tachyon $k^2=1/4$.  The world-sheet  fields have been
extended to the entire complex plane.  That is,  we have replaced 
\begin{displaymath}
\tilde{X}^{\mu}(\bar{z}) \rightarrow D^{\mu}_{\nu}X^{\nu}(\bar{z}) \ ,
\spa
\tilde{\psi}^{\mu}(\bar{z}) \rightarrow
D^{\mu}_{\nu}\psi^{\nu}(\bar{z}) \ ,
\spa
\tilde{\phi}(\bar{z}) \rightarrow \phi(\bar{z})\,, \mand
\tilde{S}_{\al}(\bar{z}) \rightarrow M_{\al}{}^{\be}{S}_{\be}(\bar{z})
 \ ,
\non\end{displaymath}
where
\begin{displaymath}
D = \left( \begin{array}{cc}
-1_{9-p} & 0 \\
0 & 1_{p+1}
\end{array}
\right) \ ,\,\, \mand 
M_p = \left\{\begin{array}{cc}\frac{\pm i}{(p+1)!}\ga^{i_{1}}\ga^{i_{2}}\ldots \ga^{i_{p+1}}
\eps_{i_{1}\ldots i_{p+1}}\,\,\,\,{\rm for\, p \,even}\\ \frac{\pm 1}{(p+1)!}\ga^{i_{1}}\ga^{i_{2}}\ldots \ga^{i_{p+1}}\ga_{11}
\eps_{i_{1}\ldots i_{p+1}} \,\,\,\,{\rm for\, p \,odd}\end{array}\right. 
\non\end{displaymath}
Using these replacements, one finds the standard propagators for the world-sheet fields $X^{\mu}\,, \phi$, \ie
\begin{eqnarray}
\lan X^{\mu}(z)X^{\nu}(w)\ran & = & -\eta^{\mu\nu}\log(z-w) \ , \non \\
\lan\phi(z)\phi(w)\ran & = & -\log(z-w) \ .
\labell{prop}\end{eqnarray}
One also needs the correlation function between two spin operators. 
The correlation function involving an arbitrary number of $\psi$'s 
and 
two $S$'s is obtained using the following Wick-like rule~\cite{Liu:2001qa}:
 \beqa
 \lan\psi^{\mu_1}
(y_1)...
\psi^{\mu_n}(y_n)S_{\al}(z)S_{\be}(\bz)\ran&\!\!\!\!=\!\!\!\!
&\frac{1}{2^{n/2}}
\frac{(z-\bz)^{n/2-5/4}}
{|y_1-z|...|y_n-z|}\left[(\Gamma^{\mu_n...\mu_1}
C^{-1})_{\al\be}\right.\nonumber\\&&+
\lan\psi^{\mu_1}(y_1)\psi^{\mu_2}(y_2)\ran(\Gamma^{\mu_n...\mu_3}
C^{-1})_{\al\be}
\pm perms\nonumber\\&&+\lan\psi^{\mu_1}(y_1)\psi^{\mu_2}(y_2)\ran
\lan\psi^{\mu_3}(y_3)\psi^{\mu_4}(y_4)\ran(\Gamma^{\mu_n...\mu_5}
C^{-1})_{\al\be}\nonumber\\&&\left.
\pm {\rm perms}+\cdots\right]\labell{wicklike}\eeqa where  dots mean  sum  over all possible contractions. In above equation, $\Gamma^{\mu_{n}...\mu_{1}}$ is the totally antisymmetric combination of the gamma matrices and  the Wick-like contraction, for  real $y_i$, is given by \beqar\lan\psi^{\mu}(y_{1})\psi^{\nu}(y_{2})\ran &=&2\eta^{\mu\nu}{\frac {Re[(y_{1}-z)(y_{2}-\bz)]}{(y_{1}-y_{2})(z-\bz)}}\eeqar 
The number of $\psi$ in the correlators  \reef{cor10} is zero. Using the above formula for zero $\psi$ and performing the other correlators using \reef{prop}, one finds that the integrand  is invariant under $SL(2,R)$ transformation. Gauge fixing this symmetry by 
 fixing  the position of vertex operators at  $(x,y,z,\bar{z}) = (x,-x,i,-i)$, one finds \cite{Kennedy:1999nn} 
 \beqa
{\cal A}^{T,RR} & \sim &  2i\Tr
(P_{-}\fsH_{(n)}M_p)\Tr(\lam) \ , \non \\
&=&\left(\frac{i\pi\beta'\mu_p'}{2}\right)
 \Tr (P_{-}\fsH_{(n)}M_p)\Tr(\lam) \labell{amp331}\ .
\eeqa
where the  conservation of momentum along the world volume of brane, \ie $k^{i} + p^{i} =0$, has been used. We have also normalized the amplitude by $\pi\beta'\mu_p'/4$.

The trace in \reef{amp331} containing the factor of $\ga^{11}$ ensures the following
results also hold for $p>3$ with $H_{(n)} \equiv \ast H_{(10-n)}$ for
$n\geq 5$. The trace is zero for $p+1\neq n$, and for $n=p+1$ it is
\begin{displaymath}
\Tr \left( \fsH_{(n)}M_p\right)
= \pm\frac{ 32}{(p+1)!}H_{i_{1}\cdots i_{p+1}}
 \eps^{i_{1}\cdots i_{p+1}} \ .
\non\end{displaymath}
We are going to compare string theory S-matrix elements with field theory S-matrix elements including their coefficients, however, we are not interested in fixing the overall sign of the amplitudes. Hence, in above and in the rest of equations in this paper, we have payed   no attention to the sign of  equations. Replacing the trace into \reef{amp331}, one finds the following coupling:
\beqa
2\beta'\mu_p' (2\pi\alpha')\Tr\left(C_{p}\wedge DT\right)
\eeqa
Note that to compare the string result with the field theory result, one has to set $\alpha'=2$ in the field theory.

\section{The $T-A-C$ amplitude}

 The three point amplitude between one RR, one  tachyon and one gauge field is given by the following correlation functions:
\begin{eqnarray}
{\cal A}^{T,A,RR} & \sim & \int dxdyd^2z
 \lan 
V_{T}^{(-1)}(x)V_A^{(0)}(y)
V_{RR}^{(-1)}(z,\bar{z})\ran\labell{cor1}\eeqa
where the tachyon and RR vertex operators are  given in \reef{vertex1} and
\beqa
V_A^{(0)}(y)&=&\xi_i(\prt X^i(y)+2ik_2\inn\psi(y)\psi^i(y))e^{2ik_2\inn X(y)}\lam\otimes I\labell{vertex3}\eeqa
The $\sigma$-factor of the above S-matrix element is the same as the $\sigma$ factor of the S-matrix element in the previous section, \ie $\Tr(\sigma_2I\sigma_3\sigma_1)=2i$. So the CP factor is $2i\Tr(\lam_1\lam_2)$.  Using \reef{prop} and \reef{wicklike},  one can perform the correlators above and show that the integrand is $SL(2,R)$ invariant. Fixing it as before, one finds
\beqa
 \int_{-\infty}^{\infty} \d x
\left( \frac{(1+x^{2})^{2}}{16 x^{2}}\right)^{1/4 + u} \frac{2}{1+x^{2}} \left(\Tr
(P_{-}\fsH_{(n)}M_p\Gamma^{ba})k_{1a}\xi_b+\frac{1-x^2}{2x}k_1\inn\xi\Tr (P_{-}\fsH_{(n)}M_p)\right)\nonumber\eeqa
where $u = -(k_1+k_2)^2$ and conservation of momentum along the world volume of brane, \ie $k_1^{i} + k_2^{i} + p^{i} =0$, has been used. While the first term satisfies the Ward identity, the second term does not. However, the integral of the second term is zero. This indicates that there is no coupling between one $C_{p}$, one tachyon and one gauge field. This is consistent with the coupling in the first line of \reef{exp2}  because $\Tr[A,T]=0$.  The integral of the first term is
\beqa
{\cal A}^{T,A,RR} &=&(\pi\beta'\mu_p')2\pi \frac{\Ga[-2u+1/2]}{\Ga[3/4-u]^2}
 \Tr
(P_{-}\fsH_{(n)}M_p\Gamma^{ba})k_{1a}\xi_b\Tr(\lam_1\lam_2) \labell{amp33}\ .
\eeqa
where we have also normalized the amplitude by the factor $ (-i\pi\beta\mu_p'/2)$.  The trace is zero for $p\neq n+1$, and for $n+1=p$ it is
\beqa
\Tr\bigg(\fsH_{(n)}M_p
(k_1.\ga)(\xi.\ga)\bigg)\delta_{p,n+1}&=&\pm\frac{32}{(p-1)!}\eps^{i_{1}\cdots i_{p+1}}H_{i_{1}\cdots i_{p-1}}
k_{1i_{p}}\xi_{i_{p+1}}\delta_{p,n+1}\nonumber\eeqa
As can be seen from the poles of the Gamma function, the above amplitude has neither  massless pole nor tachyon pole. This is consistent with the WZ terms in \reef{exp2}. However, there are infinite higher derivative couplings between one $C_{p-2}$, one tachyon and one gauge field which should be higher derivative extension of the WZ coupling $C_{p-2}\wedge F\wedge DT$. They can be read from the momentum expansion of the above S-matrix element.

The momentum expansion should be  either around $(k_1+k_2)^2\rightarrow 0$ or around $k_1\inn k_2\rightarrow 0$. The momentum expansion can be found   by studying  the massless/tachyon pole of field theory \cite{Garousi:2002wq}. 
Since there is no massless/tachyon pole,  one may conclude that the  momentum expansion of this amplitude can not be found. However, the tachyon pole of the Feynman amplitude disappear because of the kinematic reason, \ie the off-shell tachyon is abelian, and the nonabelain kinetic term of tachyon in which one of the tachyons is abelian is zero. So the momentum expansion should be  around $u\rightarrow -1/4$, or in terms of momentum around $k_1\inn k_2\rightarrow 0$. Note that  the on-shell condition implies the RR field must be non-zero, \ie $p_ip^i\rightarrow 1/4$ which is possible only for non-BPS SD-brane.
Expansion of the prefactor at this point is
\beqa
2\pi\frac{\Ga[-2u+1/2]}{\Ga[3/4-u]^2}
 &=&    2\pi\sum_{n=-1}^{\infty}b_n(u+1/4)^{n+1} 
\ .\labell{taylor1}\nonumber
\eeqa
where some of the coefficients $b_n$ are
\beqa
b_{-1}&=&1\nonumber\\
b_0&=&0\nonumber\\
b_1&=&\frac{\pi^2}{6}\nonumber\\
b_2&=&2\z(3)\nonumber\\
b_3&=&\frac{19}{360}\pi^4\labell{bs}\\
b_4&=&\frac{1}{3}\left(\z(3)\pi^2+18\z(5)\right)\nonumber\\
b_5&=&\frac{1}{3024}\left(55\pi^6+6048\z(3)^2\right)\nonumber\eeqa
We will see in the next section that the whole infinite contact terms of the above  momentum expansion  appear in the massless pole of the S-matrix element of one $C_{p-2}$ and three tachyons. Note that the  above coefficients  are exactly the same as the coefficients appear in the momentum expansion of the S-matrix element of one RR and two gauge fields \cite{Garousi:2007si}. Replacing the above expansion in \reef{amp33}, one finds  that the higher derivative extension of the WZ coupling $C_{p-2}\wedge F\wedge DT$ is the following:
\beqa
2\beta'\mu_p'(2\pi\alpha')^2C_{p-2}\wedge \Tr\left(\sum_{n=-1}^{\infty}b_n(\alpha')^{n+1}D^{a_1}\cdots D^{a_{n+1}} F\wedge D_{a_1}\cdots D_{a_{n+1}} DT\right)\labell{highaa}\eeqa
which is the same as  the higher derivative extension of the WZ coupling $C_{p-3}\wedge F\wedge F$ of the brane-anti-brane system.  This may indicate that  the higher derivative WZ couplings  of  non-BPS brane and  the higher derivative  WZ couplings of  brane-anti-brane have the same structure. In the next section we calculate the S-matrix element of one RR and three tachyons.


\section{The $T-T-T-C$ amplitude}
The S-matrix element of  one RR field and three   tachyons  is given by the following correlation function:
\begin{eqnarray}
{\cal A}^{TTTC} & \sim & \int dx_{1}dx_{2}dx_{3}dzd\bar{z}\,
  \lan V_T^{(-1)}{(x_{1})}V_{T}^{(0)}{(x_{2})}
V_{T}^{(0)}{(x_{3})}
V_{RR}^{(-1)}(z,\bar{z})\ran\nonumber\eeqa
where we have chosen the vertex operators according to the rule that the
total superghost number must be $-2$ . The tachyon in -1 picture and RR vertex operators are given in \reef{vertex1}. The tachyon in 0 picture is 
\beqa
V_T^{(0)}(x)&=&2ik\inn\psi(x)e^{2ik\inn X(x)}\lam\otimes\sigma_1
\eeqa
  
Introducing $x_{4}\equiv\ z=x+iy$ and $x_{5}\equiv\bz=x-iy$,  the scattering amplitude reduces to the following 
 correlators:\beqa {\cal A}^{TTTC}&\sim& \int 
 dx_{1}\cdots dx_{5}\,
(P_{-}\fsH_{(n)}M_p)^{\al\be}<:e^{-1/2\phi(x_4)}:
e^{-1/2\phi(x_5)}:e^{-\phi(x_1)}:>\labell{atttc} \\
&&\times {<:e^{2ik_1\cdot X(x_1)}:e^{2ik_2\cdot X(x_2)}
:e^{2ik_3\cdot X(x_3)}:e^{ip\cdot X(x_4)}:e^{ip\cdot D\cdot X(x_5)}:>}
\  \non \\&&\times {<:S_{\al}(x_4):S_{\be}(x_5)
:2ik_2.{\psi(x_2)}:2ik_3.{\psi(x_3)}:>}\Tr(\lam_1\lam_2\lam_3)\Tr(\sigma_2\sigma_1\sigma_1\sigma_3\sigma_1)\nonumber\eeqa
 The $\sigma$-factor in the above amplitude is $\Tr(\sigma_2\sigma_1\sigma_1\sigma_3\sigma_1)=2i$ for any permutation of the tachyons.

The correlators in the first and the second lines can be calculated using the propagators in \reef{prop}, and the correlator in the last line can be read from \reef{wicklike}. The result is 
\begin{eqnarray}
&&4i\Tr(\lam_1\lam_2\lam_3)\int dx_{1}\cdots dx_{5}\,k_{2i}k_{3j}
(x_{34}x_{35})^{-1/2+2k_{3}.p}{(x_{14}x_{15})^{-1/2+2k_{1}.p}}
(x_{24}x_{25})^{-1/2+2k_{2}.p}
\nonumber\\
&& {x_{45}^{p.D.p-1/2}}{{x_{23}^{4k_2.k_3}x_{12}^
{4k_1.k_2}}{x_{13}^{4k_1.k_3}}}(P_{-}\fsH_{(n)}M_p)^{\al\be}
\bigg\{(\Gamma^{ij}C^{-1})_{\al\be}+
 2{\h^{ij}}{\frac{Re[x_{24}x_{35}]}{x_{23}x_{45}}}(C^{-1})
 _{\al\be}\bigg\}\non\end{eqnarray}
where ${x_{ij}}\equiv\ x_{i}-x_{j}$. One can show that the integrand is invariant under
SL(2,R) transformation. Fixing   this  symmetry as \beqar
 x_{1}&=&0 ,\qquad x_{2}=1,\qquad x_{3}\rightarrow \infty,
 \qquad dx_1dx_2dx_3\rightarrow x_3^{2}
 \eeqar One  finds 
 \beqa {\cal A}^{TTTC}&\sim& 4i\Tr(\lam_1\lam_2\lam_3)\int
dx_4dx_5\,k_{2i}k_{3j}{x_{45}^
{-2(t+s+u)-2}}{|x_{4}|^{2t+2s}}{|1-x_{4}|^{2u+2t}}
\nonumber\\&&\times
 \bigg\{(\Gamma^{ij}C^{-1})_{\al\be}
+2{\h^{ij}}{\frac{x-1}{x_{45}}}(C^{-1})_{\al\be}\bigg\}
(P_{-}\fsH_{(n)}M_p)^
 {\al\be}\labell{extra}\eeqa 
where we have also introduced   the Mandelstam variables \beqa
s&=&-(k_1+k_3)^2,\qquad t=-(k_1+k_2)^2,\qquad u=-(k_2+k_3)^2
\qquad\labell{man}\eeqa
 and used the conservation of momentum along the brane, \ie $k_{1}^{i}+k_{2}^{i}
+k_{3}^{i}+p^{i}=0$. They satisfy the on-shell condition 
\beqa
s+t+u&=&-3/4-p_ip^i\non\eeqa 
Using the fact that $M_p$, $\fsH_{(n)}$, and $\Gamma^{ji}$ are totally antisymmetric combinations of the Gamma matrices, one realizes that the first term is non-zero only for $p=n+1$, and the last  term is  non-zero only for $p=n-1$. The integral in the above equation can be written in terms of the Gamma functions, using the following identity \cite{Fotopoulos:2001pt}:
\beqa 
 \int d^2 \!z |1-z|^{a} |z|^{b} (z - \bar{z})^{c}
(z + \bar{z})^{d}&\!\!\!\!=\!\!\!&
(2 i)^{c} 2^d \,  \pi \frac{ \Gamma( 1+ d +
\frac{b+c}{2})\Gamma( 1+ \frac{a+c}{2})\Gamma( -1-
\frac{a+b+c}{2})\Gamma( \frac{1+c}{2})}{
\Gamma(-\frac{a}{2})\Gamma(-\frac{b}{2})\Gamma(2+c+d+
\frac{a+b}{2})}\nonumber
\eeqa
for $ d= 0,1$ and arbitrary $a,b,c$. The region of integration is the upper half complex plane, as in our case. Using this integral, one finds that ${\cal A}^{TTTC}$ is equal to 
\beqa
\frac{3\beta'\mu_p'}{2\sqrt{\pi}}\Tr(\lam_1\lam_2\lam_3)\left[\Tr\bigg(P_{-}\fsH_{(n)}M_p
(k_2.\ga)(k_3.\ga)\bigg)I\delta_{p,n+1}
-i\Tr\bigg(P_{-}\fsH_{(n)}M_p
\bigg)J\delta_{p,n-1}\right]\labell{gen}\nonumber\eeqa
where we have normalized the amplitude by $-i3\beta'\mu_p'/8\sqrt{\pi}$. There are similar terms with coefficient $\Tr(\lam_1\lam_3\lam_2)$. The extra factor of $i$ in the second term is coming from the fact that the second term in \reef{extra} has factor $x_{45}=2iy$. In above equation,  $I,\,J$ are :
 \beqa I&=&(2)^{-2(t+s+u)-1}\pi{\frac{\Gamma(-u)
 \Gamma(-s)\Gamma(-t)\Gamma(-t-s-u-1/2)}{\Gamma(-u-t)
\Gamma(-t-s)\Gamma(-s-u)}}\nonumber\eeqa
\beqa J&=&(2)^{-2(t+s+u)-2}\pi{\frac{\Gamma(-u+1/2)
\Gamma(-s+1/2)\Gamma(-t+1/2)\Gamma(-t-s-u-1)}
{\Gamma(-u-t)\Gamma(-t-s)\Gamma(-s-u)}}\nonumber\eeqa 
The  traces are: \beqa
\Tr\bigg(\fsH_{(n)}M_p
(k_2.\ga)(k_3.\ga)\bigg)\delta_{p,n+1}&=&\pm\frac{32}{n!}\eps^{i_{1}\cdots i_{p+1}}H_{i_{1}\cdots i_{p-1}}
k_{2i_{p}}k_{3i_{p+1}}\delta_{p,n+1}\nonumber\\
\Tr\bigg(\fsH_{(n)}M_p
\bigg)\delta_{p,n-1}&=&\pm\frac{32}{n!}\eps^{i_{1}\cdots i_{p+1}}H_{i_{1}\cdots i_{p+1}}
\delta_{p,n-1}\nonumber
\eeqa
Note that the amplitude is symmetric under interchanging $s,t,u$. Examining  the  the Feynman diagrams resulting from the WZ couplings in \reef{exp2}, one realizes   that for the case that $p=n+1$ there are massless poles in $s-$, $t-$ and $u-$channels, and for the case that $p=n-1$ there is tachyon pole in $(s+t+u)-$channel. So the expansion should be around the following points:
\beqa
\frac{1}{3}\left([u\rightarrow 0,\, s,t\rightarrow -1/2]+[t\rightarrow 0,\, s,u\rightarrow -1/2]+[s\rightarrow 0,\, u,t\rightarrow -1/2]\right)\labell{ex}\eeqa
To see that it is momentum expansion, we write it in terms of momenta of the tachyons, \ie 
\beqa
&&\frac{1}{3}\left([(k_2+k_3)^2,k_1\inn k_3,k_1\inn k_2\rightarrow 0]+[(k_1+k_2)^2,k_1\inn k_3,k_2\inn k_3\rightarrow 0]\right.\nonumber\\
&&\left.\qquad\qquad\qquad+[(k_1+k_3)^2,k_1\inn k_2,k_2\inn k_3\rightarrow 0]\labell{ex1}\right)\eeqa
so the expansion should be a momentum expansion.  Note again that the on-shell condition implies that the RR field must be non-zero, \ie $p_ip^i\rightarrow 1/4$ which is possible only for non-BPS SD-branes. Let us study each case separately. 

\subsection{$p=n+1$ case}

 The electric part of the amplitude for one $C_{p-2}$ and three tachyons is  
\beqa
{\cal A}^{TTTC}&=&\pm\frac{24\beta'\mu_p'}{\sqrt{\pi}(p-1)!}\left[ \eps^{i_{1}\cdots i_{p+1}}H_{i_{1}\cdots i_{p-1}}
k_{2i_{p}}k_{3i_{p+1}}\right]I\,\Tr(\lam_1\lam_2\lam_3)\labell{pn2}\eeqa 
Note that apart from the group factor the amplitude is antisymmetric under interchanging the tachyons. So the whole amplitude is zero for abelian gauge group. To find the couplings in non-abelian theory, we expand $I$ around \reef{ex}, \ie
\beqa
I&=&
\frac{2\pi\sqrt{\pi}}{3}\left(-\frac{1}{u}\sum_{n=-1}^{\infty}b_n(s+t+1)^{n+1}
+\sum_{p,n,m=0}^{\infty}c_{p,n,m}u^p\left((s+1/2)(t+1/2)\right)^n(s+t+1)^m\right.\nonumber\\
&-&\left.\frac{1}{t}\sum_{n=-1}^{\infty}b_n(s+u+1)^{n+1}
+\sum_{p,n,m=0}^{\infty}c_{p,n,m}t^p\left((s+1/2)(u+1/2)\right)^n(s+u+1)^m\right.\nonumber\\
&-&\left.\frac{1}{s}\sum_{n=-1}^{\infty}b_n(u+t+1)^{n+1}
+\sum_{p,n,m=0}^{\infty}c_{p,n,m}s^p\left((u+1/2)(t+1/2)\right)^n(u+t+1)^m\right)\nonumber\eeqa
where the coefficients $b_n$ are exactly those that appear in \reef{bs} and  $c_{p,0,0}=a_p$ are 
\beqa
a_0&=&4\ln(2)\nonumber\\
a_1&=&\frac{\pi^2}{6}-8\ln(2)^2\nonumber\\
a_2&=&-\frac{2}{3}\left(\pi^2\ln(2)-3\z(3)-16\ln(2)^3\right)\labell{as}\\
a_3&=&\frac{1}{120}\left(160\pi^2\ln(2)^2+3\pi^4-960\z(3)\ln(2)-1280\ln(2)^4\right)\nonumber\\
a_5&=&-\frac{1}{90}\left(160\pi^2\ln(2)^3+9\pi^4\ln(2)-1440\z(3)\ln(2)^2+30\z(3)\pi^2-768\ln(2)^5-540\z(5)\right)\nonumber
\eeqa
 The  constants $c_{p,n,m}$ for some other cases are the following:
\beqa
&&c_{0,0,2}=\frac{2}{3}\pi^2\ln(2),\,c_{0,1,0}=-14\z(3),
c_{0,0,3}=8\z(3)\ln(2),\,\nonumber\\
&&c_{1,1,0}=56\z(3)\ln(2)-\pi^4/2,\,c_{1,0,2}=\frac{1}{36}(\pi^4-48\pi^2\ln(2)^2),\,c_{0,1,1}=-\pi^4/2\nonumber
\eeqa
It is interesting to note that the above coefficients are exactly those that appear in the momentum expansion of the S-matrix element of one $C_{p-3}$, two tachyons and one gauge field of brane-anti-brane system \cite{Garousi:2007si}. 
 This may indicate that the higher derivative couplings of non-BPS brane and  brane-anti-brane have  the same structure. The  massless poles and the contact terms of the S-matrix element \reef{pn2}   correspond to the  Feynman diagrams in Fig.1.
\begin{center}
\begin{picture}
(600,100)(0,0)
\Line(25,105)(75,70)\Text(50,105)[]{$T$}
\Line(25,35)(75,70)\Text(50,39)[]{$T$}
\Photon(75,70)(125,70){4}{7.5}\Text(105,88)[]{$A$}
\Gluon(125,70)(175,105){4.20}{5}\Text(145,105)[]{$C_{p-2}$}
\Line(125,70)(175,35)\Text(150,39)[]{$T$}
\SetColor{Black}
\Vertex(75,70){1.5} \Vertex(125,70){1.5}
\Text(105,20)[]{(a)}
\Line(295,105)(345,70)\Text(320,105)[]{$T$}
\Line(295,70)(345,70)\Text(315,80)[]{$T$}
\Line(295,35)(345,70)\Text(315,35)[]{$T$}
\Gluon(345,70)(395,105){4.20}{5}\Text(365,105)[]{$C_{p-2}$}
\Vertex(345,70){1.5}
\Text(345,20)[]{(b)}

\end{picture}\\ {\sl Figure 1 : a) The Feynman diagram corresponding to the amplitude \reef{amp3}, b) the Feynman diagram corresponding to the couplings \reef{hderv12}.} 
\end{center}

 The contact terms in \reef{pn2} can be reproduced by the following couplings:
\beqa
&&8i\beta'\alpha'(\pi\alpha')\mu_p'\sum_{p,n,m=0}^{\infty}c_{p,n,m}\left(\frac{\alpha'}{2}\right)^{p}\left(\alpha'\right)^{2n+m} C_{p-2}\wedge \Tr\left(\frac{}{}D^{a_1}\cdots D^{a_{2n}} D^{b_1}\cdots D^{b_{m}}DT\right.\nonumber\\
&&\left.\wedge(D^aD_a)^p  D_{b_1}\cdots D_{b_{m}}(D_{a_1}\cdots D_{a_n}DT\wedge D_{a_{n+1}}\cdots D_{a_{2n}}DT)\frac{}{}\right)\labell{hderv12}
\eeqa
which is the higher derivative extension of $C_{p-2}\wedge DT\wedge DT\wedge DT$. This fixes the normalization constant $\beta'$ to be 
\beqa
\beta'&=&\frac{1}{\pi}\sqrt{\frac{6\ln(2)}{\alpha'}}\labell{norm2}\eeqa
Note that the normalization of the tachyon in the WZ terms of non-BPS brane is different from the normalization of tachyon in the WZ terms of brane-anti-brane \reef{norm0}. The above higher derivative extension is  the same as the higher derivative extension of the coupling $C_{p-3}\wedge F\wedge DT\wedge DT^*$ of brane-anti-brane system.

The  field theory has  
 the following massless pole for $p=n+1$:
\beqa
{\cal A}&=&V_a^{\alpha}(C_{p-2},T,A)G_{ab}^{\alpha\beta}(A)V_b^{\beta}(A,T,T)\labell{amp3}\eeqa
where the vertices can be found from the standard nonabelian kinetic term of tachyon and from the higher derivative couplings in \reef{highaa}, \ie\footnote{Note that the coupling of two tachyons and one gauge field is given by the S-matrix element of two tachyons in -1 picture and one gauge field in 0 picture. Hence the internal CP factor is $\Tr(I\sigma_2\sigma_2)=2$ for the two permutation of the tachyons. One should not consider the case that the gauge field and one of the tachyon to be in  -1 picture and the other tachyon to be in 0 picture, because the CP factor is $\Tr(\sigma_3\sigma_2\sigma_1)=-2i$ in one ordering and is $\Tr(\sigma_3\sigma_1\sigma_2)=2i$ in other ordering of the tachyons. So the latter choice does not produce the nonabelian kinetic term of tachyon.}
\beqa
G^{\alpha\beta}_{ab}(A) &=&\frac{i\delta_{ab}\delta_{\alpha\beta}}{(2\pi\alpha')^2 T_p
\left(u\right)}\nonumber\\
V^{\beta}_b(A,T,T)&=&iT_p(2\pi\alpha')(k_2-k_3)_b[\Tr(\lam_2\lam_3\Lambda^{\beta})-\Tr(\lam_3\lam_2\Lambda^{\beta})]\nonumber\\
V^{\alpha}_a(C_{p-2},T,A)&=&2\beta'\mu_p'(2\pi\alpha')^2\frac{1}{(p-1)!}\epsilon_{i_1\cdots i_{p}a}H^{i_1\cdots i_{p-1}}k_1^{i_{p}}\sum_{n=-1}^{\infty}b_n(\alpha'k_1\cdot k)^{n+1}\Tr(\lam_1\Lambda^{\alpha})\nonumber\eeqa
where $k$ is the momentum of the off-shell gauge field. The amplitude \reef{amp3} becomes
\beqa
{\cal A}&=&2\beta'\mu_p'(2\pi\alpha')\frac{2}{(p-1)!u}\epsilon_{i_1\cdots i_{p+1}}H^{i_1\cdots i_{p-1}}k_2^{i_{p}}k_3^{i_{p+1}}[\Tr(\lam_1\lam_2\lam_3)-\Tr(\lam_1\lam_3\lam_2)]\nonumber\\
&&\times\sum_{n=-1}^{\infty}b_n\left(\frac{\alpha'}{2}\right)^{n+1}(s+t+1)^{n+1}\labell{AA}\nonumber\eeqa
this is exactly the massless pole of the string theory amplitude \reef{pn2}. Note that there is no left over residual contact term in comparing above amplitude with the massless pole of the string theory amplitude. 

\subsection{$p=n-1$ case}

Now we consider  $p=n-1$ case.  The electric part is,
\beqa
{\cal A}^{TTTC}&=&\pm\frac{24i\beta'\mu_p'}{\sqrt{\pi}(p+1)!}\bigg( \eps^{i_{1}\cdots i_{p+1}}H_{i_{1}\cdots i_{p+1}}\bigg)J\,\Tr(\lam_1\lam_2\lam_3)\labell{pn}\eeqa
The amplitude is symmetric under interchanging the tachyons. So it is non-zero even for abelian case. The expansion of $J$  around \reef{ex} is
\beqa
&&J=\sqrt{\pi}\left(\frac{-1}{(t+s+u+1)}+ \sum_{n=0}^{\infty}a_n(s+t+u+1)^n\right.\nonumber\\
&&\left.+\frac{\sum_{n,m=0}^{\infty}d_{n,m}[(s+t+1)^n((t+1/2)(s+1/2))^{m+1}+(t,s\rightarrow t,u)+(t,s\rightarrow s,u)]}{3(t+s+u+1)}\right.\labell{high}\\
&&\left.+\sum_{p,n,m=0}^{\infty}\frac{e_{p,n,m}}{3}(s+t+u+1)^p\right.\nonumber\\
&&\left.\qquad\qquad\qquad\times\left[(s+t+1)^n((t+1/2)(s+1/2))^{m+1}+(t,s\rightarrow t,u)+(t,s\rightarrow s,u)\right]\right)\nonumber\eeqa
where the coefficients $a_n$ in the first line are exactly those appear in \reef{as}. Some of the coefficients $d_{n,m}$ and $e_{p,n,m}$ are\beqa
d_{0,0}=-\pi^2/3,\,&&d_{1,0}=8\z(3)\nonumber\\
d_{2,0}=-7\pi^4/45,\, d_{0,1}=\pi^4/45,\,&&\,d_{3,0}=32\z(5),\, d_{1,1}=-32\z(5)+8\z(3)\pi^2/3\nonumber\\
e_{0,0,0}=\frac{2}{3}\left(2\pi^2\ln(2)-21\z(3)\right),&& e_{1,0,0}=\frac{1}{9}\left(4\pi^4-504\z(3)\ln(2)+24\pi^2\ln(2)^2\right)\nonumber
\eeqa
Note that the contact terms in the last line of \reef{high} do not have the structure of   the contact terms in the first line of \reef{high}. They correspond to different couplings in field theory. Here also the above momentum expansion is the same as the momentum expansion of the S-matrix element of one $C_{p-1}$, two tachyons and one gauge field of the brane-anti-brane system \cite{Garousi:2007si}. The  tachyon poles and the contact terms  correspond to the  Feynman diagrams in Fig.2.
\begin{center}
\begin{picture}
(600,100)(0,0)
\Line(25,105)(75,70)\Text(50,105)[]{$T$}
\Line(25,70)(75,70)\Text(45,80)[]{$T$}
\Line(25,35)(75,70)\Text(50,39)[]{$T$}
\Line(75,70)(125,70)\Text(105,88)[]{$T$}
\Gluon(125,70)(175,105){4.20}{5}\Text(145,105)[]{$C_{p}$}
\SetColor{Black}
\Vertex(75,70){1.5} \Vertex(125,70){1.5}
\Text(105,20)[]{(a)}
\Line(295,105)(345,70)\Text(320,105)[]{$T$}
\Line(295,70)(345,70)\Text(320,80)[]{$T$}
\Line(295,35)(345,70)\Text(320,35)[]{$T$}
\Gluon(345,70)(395,105){4.20}{5}\Text(365,105)[]{$C_{p}$}
\Vertex(345,70){1.5}
\Text(345,20)[]{(b)}

\end{picture}\\ {\sl Figure 2 : a) The Feynman diagram corresponding to the amplitude  \reef{amp4}, b) the Feynman diagram corresponding to the couplings \reef{hderv2} and \reef{highpn'}.} 
\end{center}
The contact terms in the second term of \reef{high} correspond to the following couplings:
\beqa
-24\beta'\alpha'\mu_p'\sum_{n=0}^{\infty}a_n\left(\frac{\alpha'}{2}\right)^n C_{p}\wedge (D^aD_a)^n[T^2 DT]\labell{hderv2}
\eeqa
which is the higher derivative extension of the WZ term $C_p\wedge DT \,T^2$. The above couplings are similar to the higher derivative extension of $C_{p-1}\wedge F\, |T|^2$ coupling in  the brane-anti-brane system.  The contact terms in the last line of \reef{high} correspond to the following couplings:
\beqa
&&-8\beta'(\alpha')^2\mu_p'
\sum_{p,n,m=0}^{\infty}e_{p,n,m}(\alpha')^{2m+n}\left(\frac{\alpha'}{2}\right)^pH_{p+1}(D_aD^a)^p \labell{highpn'}\\
&& \left[D_b D_c D^{a_1}\cdots D^{a_n}D_{b_1}\cdots D_{b_{2m}}TD_{a_1}\cdots D_{a_n}(D^bD^{b_1}\cdots D^{b_m} TD^c D^{b_{m+1}}\cdots D^{b_{2m}}T) \right]\nonumber\eeqa
Similar higher derivative couplings with exactly the same coefficients $e_{p,n,m}$ appear in the brane-anti-brane system \cite{Garousi:2007si}.

The tachyon poles in  \reef{high} should be reproduced by the following amplitude:
\beqa
{\cal A}&=&V^{\alpha}(C_{p},T)G^{\alpha\beta}(T)V^{\beta}(T,T,T,T)\labell{amp4}\eeqa
 The propagator and vertex  $ V(C_{p},T)$ are
\beqa
G^{\alpha\beta}(T) &=&\frac{i\delta_{\alpha\beta}}{(2\pi\alpha') T_p
\left(-k^2-m^2\right)}\nonumber\\
V^{\alpha}(C_{p},T)&=&2i\beta'\mu_p'(2\pi\alpha')\frac{1}{(p+1)!}\epsilon_{i_1\cdots i_{p+1}}H^{i_1\cdots i_{p+1}}\Tr(\Lambda^{\alpha})\labell{Fey}\nonumber
\eeqa
To find the vertex $V^{\beta}(T,T,T,T)$ we need the  couplings of four tachyons. The higher derivative couplings of four tachyons in which all tachyons carry $\sigma_1$ factor has been found in\cite{Garousi:2008xp} by expanding the S-matrix element of four tachyons.    Using these higher derivative couplings of four tachyons, one can not produce the tachyon pole in \reef{high}. In particular there is no tachyon pole corresponding  the first term in \reef{high}, and the coefficients of the infinite tachyon poles in the second line of \reef{high} are not proportional to  $d_{n,m}$.  In the next section, we   find the four-tachyon couplings from the S-matrix element of four tachyons in which two tachyons are in 0 picture and the other two are in -1 picture.

\subsubsection{Four tachyon couplings} 
The four tachyon couplings in the higher derivative field theory can be found from a momentum expansion of the   S-matrix element of four tachyons in which two of them are in 0 picture and the other two are in -1 picture. There is freedom to choose  each vertex operator to be in 0 picture or in -1 picture. The result   is given by ( see \eg \cite{Garousi:2002wq}) \beqa {\cal A}&=&-12iT_p\left(A\frac{\Ga(-2t)\Ga(-2s)}{\Ga(-1-2t-2s)}+
B\frac{\Ga(-2s)\Ga(-2u)}{\Ga(-1-2s-2u)}+
C\frac{\Ga(-2t)\Ga(-2u)}{\Ga(-1-2t-2u)}\right) \nonumber\eeqa
where the Mandelstam  variables are \beqa
s&=&-\alpha'(k_1+k_2)^2/2\,\,,\nonumber\\
t&=&-\alpha'(k_2+k_3)^2/2\,\,,\nonumber\\
u&=&-\alpha'(k_1+k_3)^2/2\,\,.\labell{mandel} \nonumber\eeqa 
 and   satisfy the constraint \beqa
s+t+u&=&-1\labell{con1}\,\,. \eeqa
Note that our convention for the Mandelstam variables in this section is different from our convention in \reef{man}. The
coefficients $A,B,C$ are the CP factors. Using the fact that the tachyon vertex operators carry both $U(N)$  and internal $\sigma$ matrices, the CP factors in this case are the following:
\beqa A&=&\frac{1}{4}\left(\frac{}{}\Tr(\lam_1\lam_2\lam_3\lam_4)\Tr(\tau_1\tau_2\tau_3\tau_4)+\Tr(\lam_1\lam_4\lam_3\lam_2)\Tr(\tau_1\tau_4\tau_3\tau_2)\right)\,\,,\nonumber\\
B&=&\frac{1}{4}\left(\frac{}{}\Tr(\lam_1\lam_3\lam_4\lam_2)\Tr(\tau_1\tau_3\tau_4\tau_2)+\Tr(\lam_1\lam_2\lam_4\lam_3)\Tr(\tau_1\tau_2\tau_4\tau_3)\right)\,\,,\nonumber\\
C&=&\frac{1}{4}\left(\frac{}{}\Tr(\lam_1\lam_4\lam_2\lam_3)\Tr(\tau_1\tau_4\tau_2\tau_3)+\Tr(\lam_1\lam_3\lam_2\lam_4)\Tr(\tau_1\tau_3\tau_2\tau_4)\right)\,\,.\labell{phase0}\nonumber\eeqa 
where $\lam$'s are the $U(N)$ matrices and each $\tau$ is either $\sigma_1$ or $\sigma_2$ such that in the trace two of the $\tau$'s should be $\sigma_1$ and the other two should be $\sigma_2$.   Following \cite{Garousi:2002wq}, to find the momentum expansion of this amplitude, one should first write it  as $A=A_s+A_t+A_u$ where \beqa A
_s&=&-4iT_p\left(A\frac{\Ga(-2s) \Ga(-2t)}{\Ga(-1-2s-2t)}+
B\frac{\Ga(-2s)\Ga(-2u)}{\Ga(-1-2s-2u)}-
C\frac{\Ga(-2t)\Ga(-2u)}{\Ga(-1-2t-2u)}\right)\nonumber\\
A _u&=&-4iT_p\left(-A\frac{\Ga(-2s)
\Ga(-2t)}{\Ga(-1-2s-2t)}+
B\frac{\Ga(-2u)\Ga(-2s)}{\Ga(-1-2u-2s)}+
C\frac{\Ga(-2u)\Ga(-2t)}{\Ga(-1-2u-2t)}\right)\nonumber\\
A_t&=&-4iT_p\left(A\frac{\Ga(-2t)
\Ga(-2s)}{\Ga(-1-2t-2s)}-
B\frac{\Ga(-2s)\Ga(-2u)}{\Ga(-1-2s-2u)}+C
\frac{\Ga(-2t)\Ga(-2u)}{\Ga(-1-2t-2u)}\right)\labell{a717}\eeqa The momentum expansion of the above amplitude  should be around \cite{Garousi:2002wq}
 \beqa s-{\rm
channel}:&&\lim_{s\rightarrow 0\,,t,u\rightarrow -1/2}A_s\nonumber\\
t-{\rm channel}:&&\lim_{t\rightarrow 0\,,s,u\rightarrow
-1/2}A_t\nonumber\\
u-{\rm channel}:&&\lim_{u\rightarrow 0\,,s,t\rightarrow
-1/2}A_u\labell{lim1}\nonumber\eeqa 
These limits  are consistent with the constraint \reef{con1}. There are s-channel, t-channel and u-channel  massless poles in $A_s$,  $A_t$ and  $A_u$, respectively.

Now we follow \cite{Garousi:2008xp} to expand the  amplitude \reef{a717} around the above points.
Using the constraint \reef{con1}, we first rewrite the amplitude in the 
 following form:
\beqa A
_s&\!\!\!=\!\!\!&16iT_pt'u'\left(A\frac{\Ga(2t'+2u') \Ga(-2t')}{\Ga(1+2u')}+
B\frac{\Ga(2t'+2u')\Ga(-2u')}{\Ga(1+2t')}+
C\frac{\Ga(-2t')\Ga(-2u')}{\Ga(1-2t'-2u')}\right)\nonumber\\
A_u&\!\!\!=\!\!\!&16iT_pt's'\left(A\frac{\Ga(-2s') \Ga(-2t')}{\Ga(1-2s'-2t')}+
B\frac{\Ga(2t'+2s')\Ga(-2s')}{\Ga(1+2t')}+
C\frac{\Ga(2t'+2s')\Ga(-2t')}{\Ga(1+2s')}\right)\nonumber\\
A_t&\!\!\!=\!\!\!&16iT_pu's'\left(A\frac{\Ga(2s'+2u') \Ga(-2s')}{\Ga(1+2u')}+
B\frac{\Ga(-2u')\Ga(-2s')}{\Ga(1-2s'-2u')}+
C\frac{\Ga(2u'+2s')\Ga(-2u')}{\Ga(1+2s')}\right)\nonumber\eeqa 
where $s'=s+1/2=-\alpha'k_1\inn k_2$, $t'=t+1/2=-\alpha'k_2\inn k_3$ and $u'=u+1/2=-\alpha'k_1\inn k_3$. Note that $s+t'+u'=0$, $u+s'+t'=0$ and $t+s'+u'=0$. In the first line above, $u'$ and $t'$ are independent, in the second line $t'$ and $s'$ are independent and in the last line $u'$ and $s'$ are independent variables. Now the field theory corresponds to expanding the above amplitude around   \beqa
s',\,t',\, u'&\rightarrow &0\nonumber\eeqa
which is a momentum expansion.  The expansion of    the amplitude around the above point is
\beqa
A_s&=&16iT_pt'u'\times\nonumber\\
&&\left(\frac{A u'+B t'+C s}{4t'u' s}+\sum_{n,m=0}^{\infty}\left[a_{n,m}(A u'^nt'^m+B t'^nu'^m)+b_{n,m}C(u'^nt'^m+ t'^nu'^m)\right]\right)\nonumber\\
A_u&=&16iT_pt's'\times\nonumber\\
&&\left(\frac{C s'+B t'+A u}{4t's' u}+\sum_{n,m=0}^{\infty}\left[a_{n,m}(Cs'^nt'^m+B t'^ns'^m)+b_{n,m}A(s'^nt'^m+ t'^ns'^m)\right]\right)\nonumber\\
A_t&=&16iT_ps'u'\times\nonumber\\
&&\left(\frac{A u'+C s'+B t}{4s'u' t}+\sum_{n,m=0}^{\infty}\left[a_{n,m}(A u'^ns'^m+C s'^nu'^m)+b_{n,m}B(u'^ns'^m+ s'^nu'^m)\right]\right)\nonumber\eeqa
Some of the coefficients $a_{n,m}$ and $b_{n,m}$ are
\beqa
&&a_{0,0}=-\frac{\pi^2}{6},\,b_{0,0}=-\frac{\pi^2}{12}\nonumber\\
&&a_{1,0}=2\z(3),\,a_{0,1}=0,\,b_{0,1}=b_{1,0}=-\z(3)\nonumber\\
&&a_{1,1}=a_{0,2}=-7\pi^4/90,\,a_{2,0}=-4\pi^4/90,\,b_{1,1}=-\pi^4/180,\,b_{0,2}=b_{2,0}=-\pi^4/45\nonumber\\
&&a_{1,2}=a_{2,1}=8\z(5)+4\pi^2\z(3)/3,\,a_{0,3}=0,\,a_{3,0}=8\z(5),\nonumber\\
&&\qquad\qquad\qquad\qquad\qquad b_{0,3}=-4\z(5),\,b_{1,2}=-8\z(5)+2\pi^2\z(3)/3\nonumber\eeqa
Similar expansion has been found in \cite{Garousi:2008xp} for the S-matrix element of two tachyons and two gauge fields, and for the S-matrix element of four $\sigma_1$-type tachyons. Now, in  the $s$-channel, two tachyons of one type convert to gauge field and then the gauge field converts to two tachyons of the other type\footnote{Note that the S-matrix element of two tachyons  and one gauge field in non-PBS brane is given by  either  $<A^{(0)}T^{(-1)}T^{(-1)}>$ where the internal CP factor is $\Tr(I\sigma_2\sigma_2)=2$ or $<A^{(-2)}T^{(0)}T^{(0)}>$ where the internal CP factor is $\Tr(I\sigma_1\sigma_1)=2$. }. The CP factor  for the s-channel then becomes $A=\alpha$,  $B=\beta$, and $C=- \gamma$ where 
\beqa \alpha&=&\frac{1}{2}\left(\frac{}{}{\rm
Tr}(\lam_1\lam_2\lam_3\lam_4)+{\rm
Tr}(\lam_1\lam_4\lam_3\lam_2)\right)\,\,,\nonumber\\
\beta&=&\frac{1}{2}\left(\frac{}{}{\rm Tr}(\lam_1\lam_3\lam_4\lam_2)+{\rm
Tr}(\lam_1\lam_2\lam_4\lam_3)\right)\,\,,\nonumber\\
\gamma&=&\frac{1}{2}\left(\frac{}{}{\rm Tr}(\lam_1\lam_4\lam_2\lam_3)+{\rm
Tr}(\lam_1\lam_3\lam_2\lam_4)\right)\,\,.\labell{phase}\nonumber\eeqa 
Similarly, in the $u$-channel $C=\gamma$, $B=\beta$ and $A=- \alpha$ , and in the $t$-channel $A=\alpha$, $C=\gamma$ and $B=-\beta$.  Therefore, the amplitudes simplify to 
\beqa
A_s&=&16iT_pt'u'\times\nonumber\\
&&\left(\frac{\alpha u'+\beta t'-\gamma s}{4t'u' s}+\sum_{n,m=0}^{\infty}\left[a_{n,m}(\alpha u'^nt'^m+\beta t'^nu'^m)-b_{n,m}\gamma(u'^nt'^m+ t'^nu'^m)\right]\right)\nonumber\\
A_u&=&16iT_pt's'\times\labell{contactstu}\\
&&\left(\frac{\gamma s'+\beta t'-\alpha u}{4t's' u}+\sum_{n,m=0}^{\infty}\left[a_{n,m}(\gamma s'^nt'^m+\beta t'^ns'^m)-b_{n,m}\alpha(s'^nt'^m+ t'^ns'^m)\right]\right)\nonumber\\
A_t&=&16iT_ps'u'\times\nonumber\\
&&\left(\frac{\alpha u'+\gamma s'-\beta t}{4s'u' t}+\sum_{n,m=0}^{\infty}\left[a_{n,m}(\alpha u'^ns'^m+\gamma s'^nu'^m)-b_{n,m}\beta(u'^ns'^m+ s'^nu'^m)\right]\right)\nonumber\eeqa
If one ignores  the internal $\sigma$ matrices, the result would be the same as above in which  all terms  have positive sign. In that case, the massless poles are reproduced by non-abelian kinetic terms. However,  the above massless poles are reproduced by the following field theory:
\beqa
-T_p\Tr\left((\pi\alpha')m^2T^2+(\pi\alpha')D_aTD^aT-(\pi\alpha')^2F_{ab}F^{ba}+T^4\right)\labell{4T}\eeqa 
Note that the $T^4$ term is absent if one ignores the $\sigma$ factors \cite{Garousi:2002wq}.

The difference between the above contact terms \reef{contactstu} and the contact terms when there is no $\sigma$ matrix is the  minus sign of $b_{n,m}$. The four tachyon couplings corresponding to the latter contact terms has been found in \cite{Garousi:2008xp}.  Hence, the four tachyon couplings corresponding to the  contact  terms \reef{contactstu} are:
\beqa
T_p(\alpha')^{2+n+m}\sum_{m,n=0}^{\infty}(\cL_{8}^{nm}+\cL_{9}^{nm}+\cL_{10}^{nm}+\cL_{11}^{nm}+\cL_{12}^{nm})\labell{lagrang11}\eeqa
where
\beqa
\cL_{8}^{nm}&\!\!\!\!\!\!=\!\!\!\!\!\!&m^4\Tr\left(\frac{}{}a_{n,m}\cD_{nm}[\cT\cT\cT\cT]- b_{n,m}\cD'_{nm}[\cT\cT\cT\cT]+h.c.\frac{}{}\right)\nonumber\\
\cL_{9}^{nm}&\!\!\!\!\!\!=\!\!\!\!\!\!&m^2\Tr\left(\frac{}{}a_{n,m}[\cD_{nm}(\cT \cT D^{\alpha}\cT D_{\alpha}\cT)+\cD_{nm}( D^{\alpha}\cT D_{\alpha}\cT\cT \cT)]\right.\nonumber\\
&&\left.-b_{n,m}[\cD'_{nm}(\cT D^{\alpha}\cT  \cT  D_{\alpha}\cT)+\cD'_{nm}( D^{\alpha}\cT  \cT D_{\alpha}\cT \cT)] +h.c.\frac{}{}\right)\nonumber\\
\cL_{10}^{nm}&\!\!\!\!\!\!=\!\!\!\!\!\!&-\Tr\left(\frac{}{}a_{n,m}\cD_{nm}[D_{\alpha}\cT D_{\beta}\cT D^{\beta}\cT D^{\alpha}\cT]-b_{n,m}\cD'_{nm}[D_{\alpha}\cT D^{\beta}\cT D_{\beta}\cT  D^{\alpha}\cT]+h.c.\frac{}{}\right)\nonumber\\
\cL_{11}^{nm}&\!\!\!\!\!\!=\!\!\!\!\!\!&-\Tr\left(\frac{}{}a_{n,m}\cD_{nm}[D_{\alpha}\cT D_{\beta}\cT D^{\alpha}\cT D^{\beta}\cT] -b_{n,m}\cD'_{nm}[D_{\beta}\cT D^{\beta}\cT  D_{\alpha}\cT D^{\alpha}\cT]+h.c. \frac{}{}\right)\nonumber\\
\cL_{12}^{nm}&\!\!\!\!\!\!=\!\!\!\!\!\!&\Tr\left(\frac{}{}a_{n,m}\cD_{nm}[D_{\alpha}\cT D^{\alpha}\cT D_{\beta}\cT D^{\beta} \cT] -b_{n,m}\cD'_{nm}[D_{\alpha}\cT D_{\beta}\cT D^{\alpha}\cT  D^{\beta}\cT]+h.c. \frac{}{}\right)\nonumber\eeqa
where the higher derivative operators $\cD_{nm}$ and $\cD'_{nm}$ are  defined as
\beqa
\cD_{nm}(EFGH)&\equiv&D_{b_1}\cdots D_{b_m}D_{a_1}\cdots D_{a_n}E  F D^{a_1}\cdots D^{a_n}GD^{b_1}\cdots D^{b_m}H\nonumber\\
\cD'_{nm}(EFGH)&\equiv&D_{b_1}\cdots D_{b_m}D_{a_1}\cdots D_{a_n}E   D^{a_1}\cdots D^{a_n}F G D^{b_1}\cdots D^{b_m}H\nonumber\eeqa
It is not difficult to check that the infinite tower of four tachyons couplings \reef{lagrang11} produce the contact terms in the string theory S-matrix element \reef{contactstu}. 

\subsubsection{The tachyon poles in field theory} 

Having found the four-tachyon couplings, one can now calculate the tachyon poles in field theory \reef{amp4}. Since the off-shell tachyon is abelian, the vertex $V^{\beta}(T,T,T,T)$ of the tachyon coupling in  \reef{4T} is  $V^{\beta}(T,T,T,T)=-12iT_p(\Tr(\lam_1\lam_2\lam_3\Lambda^{\beta})+\Tr(\lam_1\lam_3\lam_2\Lambda^{\beta}))$. Replacing it in \reef{amp4}, one finds
\beqa
24i\beta'\mu_p'\frac{1}{(p+1)!}\epsilon_{a_0\cdots a_{p}}H^{a_0\cdots a_{p}}\frac{\Tr(\lam_1\lam_2\lam_3)+\Tr(\lam_1\lam_3\lam_2)}{s+t+u+1}\labell{Fey1}\nonumber
\eeqa
which is exactly the first term in \reef{high}. The Mandelstam variables are those defined in \reef{man}. The higher derivative vertex $V^{\beta}(T,T,T,T)$  can be found from the higher derivative couplings \reef{lagrang11}. The result is
\beqa
&&2iT_p
(-\alpha')^{n+m}(a_{n,m}-b_{n,m})\Tr(\lam_1\lam_2\lam_3\Lambda^{\beta})\left[\frac{}{}(s+1/2)(t+1/2)\right.\nonumber\\
&&\left.\times\left(\frac{}{}(k_3\inn k)^m(k_3\inn k_1)^n+(k_3\inn k)^n(k_3\inn k_1)^m+(k_2\inn k)^m(k_2\inn k_1)^n+
(k_2\inn k)^n(k_2\inn k_1)^m\right.\right.\nonumber\\
&&\left.\left.+(k_3\inn k)^m(k_2\inn k)^n+(k_3\inn k)^n(k_2\inn k)^m+(k_1\inn k_2)^m(k_1\inn k_3)^n+
(k_1\inn k_2)^n(k_1\inn k_3)^m\frac{}{}\right)\right.\nonumber\\
&&\left.+(1,2,3)\rightarrow (2,1,3)+(1,2,3)\rightarrow (3,2,1)\frac{}{}\right]\nonumber\eeqa
 where  $k^a$ is the momentum of the off-shell gauge field. There are similar  terms which have coefficient $ \Tr(\lam_1\lam_3\lam_2\Lambda^{\beta})$. 
Now one can write  $k_2\inn k=k_1\inn k_3-(k^2+m^2)/2$ and $k_3\inn k=k_1\inn k_2-(k^2+m^2)/2$. The $k^2+m^2$ in the above vertex will be canceled with the $k^2+m^2$ in the denominator of the tachyon propagator resulting  a bunch of contact terms of one RR and three tachyons, \ie the diagram (b) in fig.1. They should be subtracted from  the contact terms that have been extracted  from the S-matrix element of one RR and three  tachyons, \ie the couplings  in \reef{highpn'}. Let us at the moment  ignore the contact terms and consider only the tachyon poles of the amplitude \reef{amp4}, \ie diagram (a) in fig.1. Replacing the above vertex in \reef{amp4}, one finds the following tachyon pole:
\beqa
&&-16i\beta'\mu_p'\frac{ \eps^{a_{0}\cdots a_{p}}H_{a_{0}\cdots a_{p}}}{(p+1)!(s'+t'+u)}\Tr(\lam_1\lam_2\lam_3)
\sum_{n,m=0}^{\infty}\left(\frac{}{}(a_{n,m}-b_{n,m})\right.\nonumber\\
&&\left.\times\left[\frac{}{}s't'(t'^ms'^n+t'^n s'^m)+s'u'(u'^ms'^n+u'^n s'^m)+u't'(t'^mu'^n+t'^n u'^m)\right]\right)\nonumber\eeqa
where $t'=t+1/2=-\alpha' k_1\inn k_2$, $u'=u+1/2=-\alpha'k_2\inn k_3$ and $s'=s+1/2=-\alpha'k_1\inn k_3$. The above amplitude can be written in the following form:
\beqa
&&-8i\beta'\mu_p'\frac{ \eps^{a_{0}\cdots a_{p}}H_{a_{0}\cdots a_{p}}}{(p+1)!(s'+t'+u)}\Tr(\lam_1\lam_2\lam_3)
\sum_{n,m=0}^{\infty}\left(\frac{}{}d_{n,m}\right.\nonumber\\
&&\left.\times\left[(t'+s')^n(t's')^{m+1}+(t'+u')^n(t'u')^{m+1}+(u'+s')^n(u's')^{m+1}\right]\right)\nonumber\eeqa
similar identity has been checked explicitly in \cite{Garousi:2008xp} for studying the massless pole of the S-matrix element of $CTTA$. The above tachyon poles  are exactly equal to the tachyon poles in the second line of \reef{high}.

Finally, let us  return to  the contact terms that the field theory amplitude \reef{amp4} produces. Using the Binomial formula, one can write the contact terms as  the following:
\beqa
&&8i\beta\mu_p'\frac{ \eps^{a_{0}\cdots a_{p}}H_{a_{0}\cdots a_{p}}}{(p+1)!}\Tr(\lam_1\lam_2\lam_3)
\sum_{n,m=0}^{\infty}(a_{n,m}-b_{n,m})\left\{\frac{}{}\right.\nonumber\\
&&s't'\left[\left(2\sum_{\ell=1}^m \pmatrix{m\cr 
\ell}(t'^{m-\ell}s'^n+s'^{m-\ell}t'^n)+2\sum_{\ell=1}^n \pmatrix{n\cr 
\ell}(t'^{n-\ell}s'^m+s'^{n-\ell}t'^m)\right)(\alpha'k^2+\alpha'm^2)^{\ell-1} \right.\nonumber\\
&&\left.+\sum_{\ell=1,j=1}^{n,m} \pmatrix{n\cr 
\ell}\pmatrix{m\cr 
j}(t'^{n-\ell}s'^{m-j}+s'^{n-\ell}t'^{m-j})(\alpha'k^2+\alpha' m^2)^{\ell+j-1}\frac{}{}\right]\nonumber\labell{masslesspole4}
\\
&&\left.+(s',t')\rightarrow (s',u')+(s',t')\rightarrow (t',u')\frac{}{}\right\}\nonumber\eeqa
There are similar couplings in the brane-anti-brane system \cite{Garousi:2008xp}. Note that the above couplings have at least four momenta. They can be rewritten in the following form:
\beqa
&&i8\beta'\mu_p'\frac{ \eps^{a_{0}\cdots a_{p}}H_{a_{0}\cdots a_{p}}}{(p+1)!}\Tr(\lam_1\lam_2\lam_3)
\sum_{p,n,m=0}^{\infty}e'_{p,n,m}(s+t+u+1)^p\nonumber\\
&&\left[(s+t+1)^n((t+1/2)(s+1/2))^{m+1}+(s,t)\rightarrow (s,u)+(s,t)\rightarrow (t,u)\right]\nonumber\eeqa
where $e'_{p,n,m}$ can be written in term of $a_{n,m}$ and $b_{n,m}$. The contact terms of one RR and three tachyons  of \reef{high} have the above structure. Hence, the coefficients $e_{p,n,m}$ in the couplings \reef{highpn'}  should be replaced by \beqa
e_{p,n,m}\rightarrow e_{p,n,m}-e'_{p,n,m}\nonumber\eeqa
This makes the higher derivative  theory  to produce     the string theory  S-matrix element. This ends our illustration of consistency between the momentum expansion of the S-matrix element of one RR and three tachyons  around \reef{ex1} and the higher derivative couplings of the field theory.

\section{Discussion}

In this paper we have calculated the  disk level S-matrix elements of $CT$, $CTA$ and $CTTT$. The on-shell conditions indicate that these S-matrix elements should be considered  in the world-volume of non-BPS SD-branes. By finding their momentum expansion we have shown that the leading order terms of the expansions are consistent with the WZ terms of non-BPS SD-brane, and the non-leading terms are reproduced by some higher derivative extension of the WZ terms, \ie equations \reef{highaa}, \reef{hderv12}, \reef{hderv2} and \reef{highpn'}. These higher derivative terms have exactly the same coefficients as the higher derivative terms of the brane-anti-brane system \cite{Garousi:2007si}. These couplings have in general  on-shell ambiguity. One way of fixing these on-shell ambiguity is to compare the result with the off-shell couplings in BSFT. It is shown in \cite{Kraus:2000nj} that the WZ couplings are exact when the RR field is constant. So to write the couplings found in \cite{Garousi:2007si} in such a way that have no on-shell ambiguity, one should write the couplings in the momentum space in terms of the Mandelstam variables which are written in terms of RR momentum. In fact, using the on-shell conditions, one can write the expansion in terms of the RR momentum, \eg the Mandelstam variables for the S-matrix element of $CTTA$ should be sent to $t\rightarrow -1/4$, $s\rightarrow -1/4$, $u\rightarrow 0$ \cite{Garousi:2007si}. Using the on-shell conditions, one can rewrite them as $(p^2+2k_i\cdot p)\rightarrow 0$. So in this form all the higher derivative corrections are zero for constant RR field.   For non-BPS SD-branes, however, the couplings are valid when $p_ip^i\rightarrow 1/4$, so one can not compare the higher derivative couplings with the BSFT couplings.

The couplings in \reef{4T} and $a_{0,0},\, b_{0,0}$ couplings of \reef{lagrang11} are not consistent with the usual  non-abelian tachyon DBI action. This is resulted from the fact that the tachyons should be in different pictures. So one should modify the tachyon DBI action to include the internal CP factors.  To this end, we consider  each field carries two matrices, the $U(N)$ matrix and  the internal Pauli matrix. The Pauli matrix for the gauge field is the $2\times 2$ identity matrix\footnote{Note that in the S-matrix element of $n$ gauge fields, two of them should be in -1 picture which have $\sigma_3$ and the other $n-2$ should be in 0 picture which have identity matrix. The trace of the internal CP factor is the same as the case that  all gauge fields have identity matrix.}  whereas for the tachyon is either $\sigma_1$ or $\sigma_2$, \ie $T^1=T\sigma_1$, $T^2=T\sigma_2$. Inspired by the non-abelian DBI action \cite{Tseytlin:1997csa,Myers:1999ps} in which there are different transverse scalar fields, one may  modify the usual tachyon DBI action to  the following form:
\beqa
S_{DBI}&=&-\frac{T_p}{2}\int
d^{p+1}\sigma \STr\left(\frac{}{}V({ T^iT^i})\sqrt{1+\frac{1}{2}[T^i,T^j][T^j,T^i])}\right.\labell{nonab} \\
&&\qquad\qquad\left.
\times\sqrt{-\det(\eta_{ab}
+2\pi\alpha'F_{ab}+2\pi\alpha'D_a{ T^i}(Q^{-1})^{ij}D_b{ T^j})} \right)\,,\nonumber\eeqa  where 
\beqa
Q^{ij}&=&I\delta^{ij}-i[T^i,T^j]\eeqa
The superscripts $i,j=1,2$ and there is no sum over $i,j$.  After expanding the square roots one should specify the $\sigma$ factor of each tachyon in the expansion. The tachyons of 2-tachyon terms have  $\sigma_2$, half of the tachyons of 4-tachyon terms have $\sigma_1$ and the other half have $\sigma_2$. For the 6-tachyon terms, two of them should  be along  $\sigma_2$ and the others to be along $\sigma_1$ direction, and so on. The trace in above equation
should be completely symmetric between all  matrices
of the form $F_{ab},D_a{ T^i}$, $[T^i,T^j]$ and individual
${ T^i}$ of the  potential $V(T^iT^i)$. The trace is  over the group matrices and over the $\sigma$ matrices. For example,  $\Tr[T^1,T^2][T^2,T^1]=8\Tr(T^4)$ where the trace on the right hand side is over the group matrices only.  The tachyon couplings $\reef{4T}$ and the $a_{0,0},\, b_{0,0}$ terms in \reef{lagrang11} are the four tachyon couplings of the above DBI action.  

The potential $V$ in above equation has the following expansion\beqa
V( T^iT^i)&=&1+\pi\alpha'm^2{ T^iT^i}+
\frac{1}{2}(\pi\alpha'm^2{ T^iT^i})^2+\cdots
\non\eeqa  where $m^2$ is the mass squared of tachyon, \ie
$m^2=-1/(2\alpha')$. This expansion is consistent with the
potential $V({ T})=e^{\pi\alpha'm^2{ T}^2}$ which is the tachyon
potential of BSFT ~\cite{Kutasov:2000aq}. In above DBI action however there is a symmetric trace over the $\sigma$ factors and tachyon potential has also another square root term. Performing the symmetric trace, one finds
\beqa
\frac{1}{2}\STr\left(V({ T^iT^i})\sqrt{1+[T^i,T^j][T^j,T^i]}\right)&=&\left(1-\frac{\pi}{2}T^2+\frac{\pi^2}{24}T^4+\cdots\right)\left(1+T^4+\cdots\right)\nonumber\eeqa
The symmetric trace and the square root term do not change the sign of each term compare to the exponential potential $V(T)$. So one expects that the tachyon condensates at $T\rightarrow \infty$ and the tachyon potential becomes  zero at that point.

Having found the higher derivative tachyon couplings in \reef{lagrang11}, one can find the effective couplings for slowly varying tachyon. Ignoring the second covariant derivative of the tachyon, the couplings in \reef{lagrang11} are reduced to the four tachyon couplings of the action \reef{nonab} plus the following terms:
\beqa
T_pm^4\alpha'^3\z(3)\Tr\left(2D_aTTD^aTT+D_aTTTD^aT+D_aTD^aTTT\right)\nonumber\eeqa
One may conclude from the above terms that the  action \reef{nonab}  is not the effective action. However, the couplings in \reef{lagrang11} have on-shell ambiguity. That is, $m^2T\sim DDT$, so the above terms may be among the higher derivative terms that should be ignored when the second derivative of tachyon is zero. The couplings \reef{lagrang11} appear in the tachyon poles of the S-matrix element of $CTTT$, however, as it is argued in \cite{Garousi:2008xp}, if one replaces $T$ with $DDT$ it does not change the tachyon poles but produces an extra contact terms. Hence, this on-shell ambiguity may be resolved by studying the S-matrix element of four tachyons and one gauge field, in which the couplings \reef{lagrang11} appear in tachyon poles as well as the contact terms of this amplitude. This S-matrix element  has been found in \cite{BitaghsirFadafan:2006cj}.

We have considered the S-matrix element in section 4.2.1 as $<T^{(0)} T^{(0)} T^{(-1)}T^{(-1)}>$. On the other hand one can write the amplitude as $<T^{(0)} T^{(0)} T^{(0)}T^{(-2)}>$ in which the internal CP factor is $\Tr(\sigma_1 \sigma_1\sigma_1\sigma_1)=2$. The amplitude in the latter case is given by  \beqa {\cal A}&=&-4iT_p\left(\alpha\frac{\Ga(-2t)\Ga(-2s)}{\Ga(-1-2t-2s)}+
\beta\frac{\Ga(-2s)\Ga(-2u)}{\Ga(-1-2s-2u)}+
\gamma\frac{\Ga(-2t)\Ga(-2u)}{\Ga(-1-2t-2u)}\right)\labell{a1} \eeqa
On expects that the amplitude in section 4.2.1 reduces to the above amplitude after performing the trace over the internal CP matrices. In fact using the observation  that in the s-channel $A=\alpha$, $B=\beta$ $C=-\gamma$ and similarly for the other channels, one finds that the amplitude in \reef{a717} simplifies to $A_s=A_t=A_u$ and equal to the above amplitude. However, the momentum expansion is different from the momentum expansion considered in \cite{Garousi:2008xp}, \ie $A_s,A_t,A_u$ here are different from  $A_s,A_t,A_u$ in \cite{Garousi:2008xp}. That is the reason that the tachyon pole in ${\cal C}TTT$ does not produce by the tachyon couplings found in \cite{Garousi:2008xp}.
The interesting point that we have considered  $<T^{(0)} T^{(0)} T^{(-1)}T^{(-1)}>$ in section 4.2.1  is that the momentum expansion of \reef{a717} is similar to the momentum expansion of the massless transverse scalars, hence, the four tachyon coupling can be written in the tachyon DBI form \reef{nonab} in which there is a symmetric trace over $U(N)$ and over the internal CP matrices.

Therefore,  there are two different expansion for the S-matrix element $<T^{(0)} T^{(0)} T^{(0)}T^{(-2)}>$. One is the expansion considered in \cite{Garousi:2008xp} which is consistent with the usual tachyon DBI action and the other one which is consistent with \reef{nonab}. Our on-shell calculation is valid  only  for non-BPS SD-branes. On the other hand, it is expected that the usual tachyon DBI  action should describe the non-BPS D-branes. So one may conclude that the DBI part of the effective action of non-BPS SD-brane and non-BPS D-branes are not equal. To check explicitly which tachyon DBI action corresponds to non-BPS D-branes,  one needs to study the S-matrix element of four tachyons and one RR field on the world volume of D-brane-anti-D-brane in which the tachyon pole is reproduced by the ${\cal C}TT$ and $TTTT$  couplings. One of the above two expansions for $TTTT$ produces the tachyon poles, hence, it fixes the tachyon action.  It would be interesting to perform this calculation.  
 
\section*{Acknowledgment}

I would like to thank A. Ghodsi for  discussion.


\end{document}